\documentclass[twocolumn,showpacs,amssymb,aps,prl,floatfix]{revtex4}

\usepackage{epsfig}

\newcommand{\bra}{\langle}
\newcommand{\ket}{\rangle}

\newcommand{\be}{\begin{equation}}
\newcommand{\ee}{\end{equation}}
\newcommand{\bea}{\begin{eqnarray}}
\newcommand{\eea}{\end{eqnarray}}
\newcommand{\bean}{\begin{eqnarray*}}
\newcommand{\eean}{\end{eqnarray*}}

\newcommand{\half} {\frac{1}{2}}

\newcommand{\eps}{\epsilon}
\newcommand{\vareps}{\varepsilon}

\newcommand{\re}{\mbox{Re\,}}
\newcommand{\im}{\mbox{Im\,}}
\newcommand{\rmR}{{\rm R}}
\newcommand{\rmI}{{\rm I}}

\newcommand{\nn}{\nonumber}

\begin{document}                                                

\title{Can stochastic quantization evade the sign problem? \\
-- the relativistic Bose gas at finite chemical potential}

\author{Gert Aarts}
\affiliation{Department of Physics, Swansea University, Swansea SA2 8PP, 
United Kingdom}

\date{October 12, 2008}

\begin{abstract} 

 A nonperturbative study of field theories with a complex action, such as 
QCD at finite baryon density, is difficult due to the sign problem. We 
show that the relativistic Bose gas at finite chemical potential has a 
sign and `Silver Blaze' problem, similar to QCD. We then apply stochastic 
quantization and complex Langevin dynamics to study this theory with 
nonperturbative lattice simulations. Independence of chemical potential at 
small and a transition to a condensed phase at large chemical potential 
are found. Lattices of size $N^4$, with $N=4,6,8,10$, are used. We show 
that the sign problem is severe, however, we find that it has no negative 
effect using this approach. This improves the prospects of applying 
stochastic quantization to QCD at nonzero density.

\end{abstract}


\maketitle


{\em Introduction --}
 Field theories with a complex action are difficult to treat 
nonperturbatively. Because the weight in the path integral $e^{-S} = 
|e^{-S}| e^{i\varphi}$ is not real, standard numerical approaches based on 
a probability interpretation and importance sampling cannot be applied. 
This has hindered progress in condensed matter and many-body theories, 
such as frustrated quantum spin systems and strongly correlated electronic 
systems away from half filling. An outstanding example is QCD at nonzero 
baryon density.
 Several methods have been 
devised to circumvent the sign problem in QCD: reweighting 
\cite{Fodor:2001au,Fodor:2002km}, Taylor series expansion 
\cite{Allton:2002zi,Gavai:2003mf}, imaginary chemical potential and 
analytic continuation \cite{deForcrand:2002ci,D'Elia:2002gd}, and the use 
of the canonical ensemble \cite{Kratochvila:2005mk,Ejiri:2008xt} and the 
density of states \cite{Fodor:2007vv}.
 In general these methods can only be applied in a limited region of the 
phase diagram (such as high temperature and small chemical potential), 
suffer from overlap problems, and/or are restricted to small volumes.
 
At vanishing temperature, so far all methods have failed. It is 
well-known what is supposed to happen: if $\mu$ is the quark chemical 
potential (so that the fermion determinant satisfies $\det M(\mu) = [\det 
M(-\mu)]^*$), one expects a transition to a condensed phase (nuclear 
matter) at a critical chemical potential $\mu_c\approx m_N/3$, where $m_N$ 
is the nucleon mass. Below $\mu_c$, physical observables should be 
completely independent of $\mu$ (at strictly zero temperature), even 
though microscopically $\mu$ explicitly enters in the Boltzmann weight. 
This exact cancelation is known as the `Silver Blaze' problem 
\cite{Cohen:2003kd} and is intimately tied to the sign problem, as has 
been demonstrated in Random Matrix Theory 
\cite{Osborn:2004rf,Osborn:2005ss,Akemann:2004dr}.
 It is an outstanding challenge to probe the QCD
phase diagram in the low-temperature region at nonzero chemical potential.
	
The sign problem in QCD arises not because of the anticommuting nature of 
fermions. Instead, it is due to the behaviour of the action under complex 
conjugation when the chemical potential is nonzero.
 Therefore, this sign problem can also be studied in bosonic field 
theories with a chemical potential coupled to a conserved charge and an 
action satisfying $S(\mu) =[S(-\mu)]^*$. These theories suffer from 
exactly the same sign and Silver Blaze problems as QCD.
 In this Letter we consider the relativistic Bose gas at finite chemical 
potential and study its Silver Blaze problem with nonperturbative lattice 
simulations employing stochastic quantization \cite{Parisi:1980ys}. 
Stochastic quantization does not rely on the interpretation of $e^{-S}$ as 
a probability weight.
 Instead the proper distribution is obtained as the
equilibrium distribution of a stochastic process, described by a Langevin
equation. In the case that the action is complex, the Langevin dynamics
is complexified \cite{Parisi:1984cs,Klauder:1985a}. See
Ref.\ \cite{Damgaard:1987rr} for a comprehensive review and Ref.\
\cite{Karsch:1985cb} for an early application to the SU(3) spin model at
finite chemical potential.
 In the past complex Langevin dynamics has been hindered by
numerical instabilities and uncertainty about convergence, see e.g.\ Ref.\
\cite{Ambjorn:1986fz}.
 Recently some of these problems were alleviated by the use of more 
refined Langevin algorithms 
\cite{Berges:2005yt,Berges:2006xc,Berges:2007nr,Aarts:2008rr}. In 
particular, in Ref.\ \cite{Aarts:2008rr} the method has been applied to 
several models at finite chemical potential. In U(1) and SU(3) one link 
models, where the sign problem is mild, excellent results have been 
obtained. For QCD with static quarks, first results on a $4^4$ lattice 
indicate a transition from a low-density confining to a high-density 
deconfining phase. Even though the sign problem appears to be severe, 
observables are under control.
 In this Letter we provide considerable evidence that stochastic
quantization evades the sign problem and is capable of handling
the Silver Blaze problem, in the context of the relativistic Bose gas at
finite chemical potential.

{\em Bose gas at finite chemical potential --}
 We consider a self-interacting complex scalar field in $d=4$ euclidean 
dimensions, with the lattice action
\bea
\nn
 S &=& \sum_x \bigg[ \left(2d+m^2\right) \phi_x^*\phi_x 
 + \lambda\left( \phi_x^*\phi_x\right)^2
\\ &&
- \sum_{\nu=1}^4\left(  \phi_x^* e^{-\mu\delta_{\nu,4}} \phi_{x+\hat\nu} 
+ \phi_{x+\hat\nu}^* e^{\mu\delta_{\nu,4}} \phi_x \right)
\bigg].
\eea
 The lattice spacing $a=1$ and we take $m^2>0$. 
 The lattice four-volume is $\Omega=N_s^3N_\tau$, where $N_s$ ($N_\tau$) 
are the number of sites in a spatial (temporal)  direction. As usual, 
chemical potential $\mu$ is introduced as an imaginary constant vector 
potential in the temporal direction. Note that $S(\mu)=[S(-\mu)]^*$. At 
zero temperature and in the thermodynamic limit, bulk physical observables 
are strictly independent of the chemical potential as long as $\mu< 
\mu_c$, with $\mu_c$ the critical chemical potential. At $\mu=\mu_c$, one 
expects a second order phase transition to the Bose condensed phase, where 
the density $\bra n\ket = (1/\Omega)\partial\ln Z/\partial \mu$ is 
nonzero. Ignoring interactions, the critical chemical potential is given 
by $|\mu_c^{0}| = 2\mbox{arcsinh}(m/2)$ (corresponding to $|\mu_c^{0}|=m$ 
in the formal continuum limit). Interactions are expected to increase this 
value \footnote{See e.g.\ Ref.\ \cite{Endres:2006xu}. In this paper a 
reformulation of this model in terms of dual lattice variables was given, 
which avoids the sign problem, and numerical results for two rather than 
three spatial dimensions were obtained.}. The exact $\mu$-independence 
when $\mu<\mu_c$, even though microscopically $\mu$ is manifestly present, 
is the Silver Blaze problem.

{\em Complex Langevin dynamics --}
 To apply stochastic quantization, we start with the Langevin equation
 \be
 \frac{\partial\phi_x(\theta)}{\partial\theta} = -\frac{\delta 
S[\phi]}{\delta\phi_x(\theta)} +\eta_x(\theta),
\ee
 where $\theta$ is the Langevin time and $\eta$ is Gaussian random noise. 
 Since the action is complex, stochastic quantization relies on a 
complexification of the fields. The original 
field is first written in terms of two real fields $\phi_{a}$ ($a=1,2$) as 
$\phi=\frac{1}{\sqrt 2}\left(\phi_1+i\phi_2\right)$.
In terms of these the action reads
\bea
 S &=& \sum_x\bigg[ \left(
 d+\frac{m^2}{2}\right) \phi_{a,x}^2
 + \frac{\lambda}{4}\left(\phi_{a,x}^2\right)^2
 - \sum_{i=1}^3 \phi_{a, x}\phi_{a, x+\hat i}
 \nn \\ && 
 -\cosh\mu\,  \phi_{a, x}\phi_{a, x+\hat 4}
 +i\sinh\mu\, \vareps_{ab}\phi_{a, x}\phi_{b, x+\hat 4}
\bigg].
\label{eq:S}
\eea
 Here the antisymmetric tensor $\vareps_{ab}$, with 
$\vareps_{12}=-\vareps_{21}=1$, 
$\vareps_{11}=\vareps_{22}=0$, is introduced and summation over repeated 
indices 
is implied. The term proportional to $\sinh \mu$ causes the action to be 
complex.

 These real fields are now complexified as
\be
\label{eq:compl}
 \phi_a\to \phi_a^\rmR +i\phi_a^\rmI  \;\;\;\;\;\;\;\;\; (a=1,2),
\ee
and the complex Langevin equations read
\bea
 \label{eqphiR}
 \phi_{a,x}^\rmR(n+1) &=& \phi_{a,x}^\rmR(n) +\eps
K_{a,x}^\rmR(n) +\sqrt{\eps}\eta_{a,x}(n),
\\
 \label{eqphiI}
 \phi_{a,x}^\rmI(n+1) &=& \phi_{a,x}^\rmI(n) +\eps
K_{a,x}^\rmI(n).
\eea
 Here Langevin time is discretized as $\theta=n\eps$, with $\eps$ the 
time step, the noise $\eta$ is real and Gaussian, with
$\bra \eta_{a,x}(n)\ket = 0$,
$\bra \eta_{a,x}(n)\eta_{b,x'}(n')\ket =
2\delta_{nn'}\delta_{ab}\delta_{xx'}$, 
and the drift terms are determined by
 \bea
 K_{a,x}^\rmR &=&  -\re \frac{\delta S}{\delta\phi_{a,x}}\Big|_{\phi_a\to
\phi_a^\rmR+i\phi_a^\rmI},
\\
 K_{a,x}^\rmI &=&  -\im \frac{\delta S}{\delta\phi_{a,x}}\Big|_{\phi_a\to
\phi_a^\rmR+i\phi_a^\rmI}.
\eea
 Explicitly, they read
\bea
 K_{a,x}^\rmR &=&
 -\left[ 2d + m^2 +\lambda\left( \phi_{b,x}^{\rmR\,2} -
 \phi_{b,x}^{\rmI\,2} \right) \right] \phi_{a,x}^\rmR
 \nn \\ &&
 +2\lambda \phi_{b,x}^\rmR\phi_{b,x}^\rmI \phi_{a,x}^\rmI
 + \sum_i\left(\phi_{a,x+\hat i}^\rmR + \phi_{a,x-\hat i}^\rmR \right)
 \nn \\ &&
 + \cosh \mu \left(\phi_{a,x+\hat 4}^\rmR + \phi_{a,x-\hat 4}^\rmR \right)
 \nn \\ &&
 + \sinh\mu \,\,\vareps_{ab}
        \left(\phi_{b,x+\hat 4}^\rmI - \phi_{b,x-\hat 4}^\rmI \right),
 \\
K_{a,x}^\rmI &=&
-\left[ 2d + m^2  +\lambda\left(
\phi_{b,x}^{\rmR\,2} - \phi_{b,x}^{\rmI\,2} \right) \right]\phi_{a,x}^\rmI
 \nn \\ &&
 -2\lambda \phi_{b,x}^\rmR\phi_{b,x}^\rmI \phi_{a,x}^\rmR
 + \sum_i\left(\phi_{a,x+\hat i}^\rmI + \phi_{a,x-\hat i}^\rmI \right)
 \nn \\ &&
 + \cosh \mu \left(\phi_{a,x+\hat 4}^\rmI + \phi_{a,x-\hat 4}^\rmI \right)
 \nn \\ &&
 - \sinh\mu \,\,\vareps_{ab}
        \left(\phi_{b,x+\hat 4}^\rmR - \phi_{b,x-\hat 4}^\rmR \right).
\eea
 Observables are written in terms of the complexified fields using the 
replacement (\ref{eq:compl}). For instance, the square of the field 
modulus is given by
 \be
\label{eqvar}
\phi^*\phi = \half\phi_a^2\to
\half\left( {\phi_a^\rmR}^2 - {\phi_a^\rmI}^2 \right)
+ i \phi_a^\rmR\phi_a^\rmI,
\ee
and the density is given by $n = \frac{1}{\Omega}\sum_x n_x$, with
\bea
  n_x &=& \left( \delta_{ab} \sinh\mu - i \vareps_{ab}\cosh\mu \right)
\phi_{a,x}\phi_{b,x+\hat 4}
\label{eq:dens2}
\\
 &\to&
\left( \delta_{ab}\sinh\mu  - i\vareps_{ab} \cosh\mu \right)
\Big(
\phi_{a,x}^\rmR\phi_{b,x+\hat 4}^\rmR 
- \phi_{a,x}^\rmI\phi_{b,x+\hat 4}^\rmI
\nn \\ &&
+ i\left[
\phi_{a,x}^\rmR\phi_{b,x+\hat 4}^\rmI + \phi_{a,x}^\rmI\phi_{b,x+\hat 
4}^\rmR
\right]\Big).
\label{eq:dens}
\eea
 All observables now have a real and imaginary part.

\begin{figure}[t]
\centerline{\epsfig{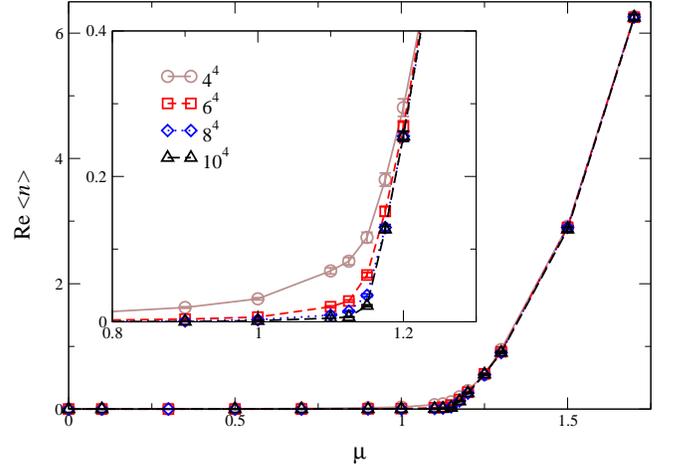}}
\caption{Real part of the density, $\re\bra n\ket$, as a function of
chemical potential for lattices of size $N^4$, with $N=4,6,8,10$. The
parameters are $m=\lambda=1$ and stepsize $\eps=5\times 10^{-5}$.
The inset shows a blowup around the transition.
In the thermodynamic limit, the density vanishes below the critical
chemical potential.
 }
\label{fig1}
\end{figure}

\begin{figure}[t] 
\centerline{\epsfig{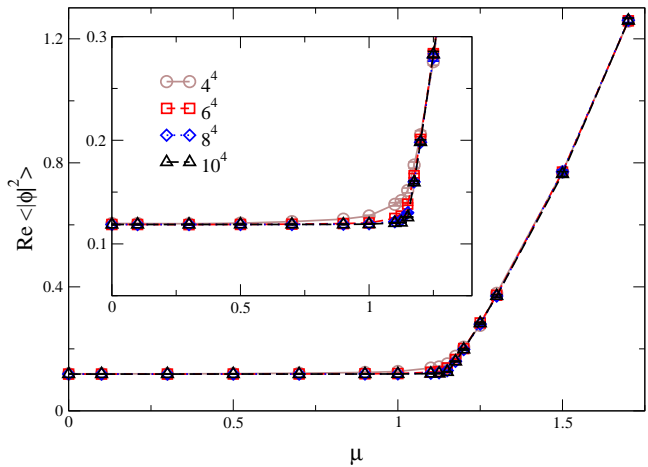}} 
\caption{Real part of the square of the field modulus, 
$\re\bra|\phi|^2\ket$, as a function of $\mu$. The inset shows a blowup at 
smaller $\mu$.
 }
\label{fig3} 
\end{figure}

{\em Results --} 
 We have solved Eqs.\ (\ref{eqphiR}, \ref{eqphiI}) numerically, using a 
Langevin stepsize $\eps=5\times 10^{-5}$, a total number of 
$5\times 10^6$ Langevin timesteps, and $m=\lambda=1$. Lattices of size 
$N^4$, with $N=4,6,8,10$ were used. No instabilities or runaway solutions 
were encountered.
 The real part of the density is shown in Fig.\ \ref{fig1} for chemical 
potentials $0\leq \mu\leq 1.7$. Imaginary parts of observables were found 
to be consistent with zero, as they should. A transition between a 
zero-density phase and a condensed phase with nonzero density is clearly 
visible. The inset shows a blowup of the transition region. Nonanalyticity 
associated with a phase transition can only occur in the thermodynamic 
limit. We observe that with increasing four-volume the transition becomes 
sharper and the density goes to zero below $\mu_c\approx 1.15$, as is 
expected in a second order phase transition \footnote{In this study we 
have not attempted to determine the precise value of $\mu_c$ or the order 
of the phase transition, which can be done using e.g.\ Binder cumulants. A 
mean field estimate of the critical chemical potential, $4\sinh^2(\mu_c/2) 
= m^2+4\lambda\bra|\phi|^2\ket$, yields $\mu_c=1.15$ 
\cite{Aarts:2009hn}.}. Finite size 
effects are comparable to what is found analytically in the noninteracting 
system, when $|\mu|<|\mu_c^0|$.
 In Fig.\ \ref{fig3} we show the real part 
of the square of the field modulus (\ref{eqvar}). Again we observe 
$\mu$-independence below $\mu_c$ and a sharp rise above. We note that the 
value at $\mu=0$ is obtained using real Langevin dynamics and is therefore 
theoretically well established. We conclude that the Silver Blaze problem 
poses no obstacle for stochastic quantization and that there is no problem 
in taking the thermodynamic limit.

Some insight into why this method works can be obtained by ignoring the 
interactions ($\lambda=0$) \cite{Aarts:2009hn}. In this case the complex 
Langevin equations can be solved analytically and convergence to the exact 
results can be proven, provided that $|\mu|<|\mu_c^{0}|$ ($=0.9624$ for 
the parameters used here). For larger $\mu$, the free theory is unstable. 
As we have shown here, interactions shift the critical chemical potential 
and remove the instability.

{\em Sign problem and phase quenching --}
 To quantify the sign problem, we write $e^{-S} = e^{-S_\rmR-iS_\rmI} = 
|e^{-S}| e^{i\varphi}$.  In reweighting, the phase factor $e^{i\varphi}$ 
is combined with the observable while simulations are performed in the 
phase-quenched theory, obtained by ignoring the imaginary part of the 
action. Expectation values in the full theory are then reconstructed as 
$\bra O\ket_{\rm full} = \bra O e^{i\varphi}\ket_{\rm pq}/\bra 
e^{i\varphi}\ket_{\rm pq}$, where the subscript pq indicates phase 
quenched. The average phase factor $\bra e^{i\varphi}\ket_{\rm pq}$ is the 
ratio of two partition functions, $\left\bra e^{i\varphi}\right\ket_{\rm 
pq} = Z_{\rm full}/Z_{\rm pq} = e^{-\Omega\Delta f}$, and vanishes 
exponentially in the thermodynamic limit $\Omega\to \infty$, when $\mu\neq 
0$. Here $\Delta f$ is the difference between the free energy densities in 
the full and phase quenched theories.

\begin{figure}[t] 
\centerline{\epsfig{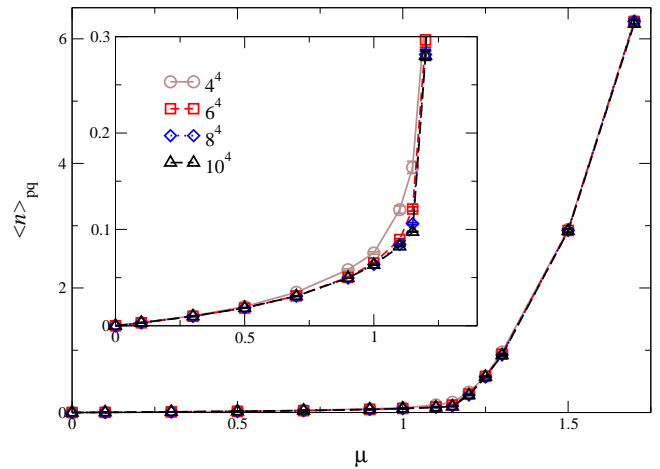}} 
\caption{Density in the phase-quenched theory, $\bra n\ket_{\rm pq}$, as 
a function of $\mu$. 
The inset shows a blowup at smaller $\mu$. 
 }
\label{fig4} 
\end{figure}

\begin{figure}[t] 
\centerline{\epsfig{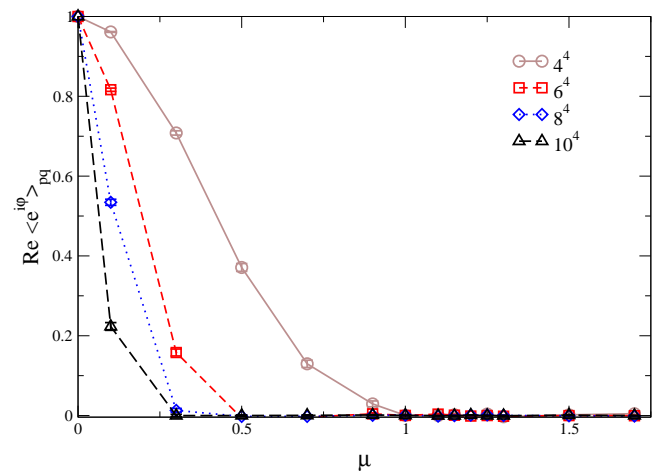}} 
\caption{Average phase factor in the phase-quenched theory, $\re\bra 
e^{i\varphi}\ket_{\rm pq}$, as a function of $\mu$, indicating the 
severeness of 
the sign problem in the thermodynamic limit. 
 }
\label{fig5} 
\end{figure}

We have studied the phase-quenched theory numerically. In this case, real 
Langevin dynamics can be employed, using only Eq.\ (\ref{eqphiR}) with 
$\phi_a^\rmI\equiv 0$. In the expression for the density (\ref{eq:dens2}), 
only the term proportional to $\sinh\mu$ is preserved. Simulations were 
carried out using the same parameters as above. No difference in 
convergence properties were found. The density in the phase-quenched 
theory is shown in Fig.\ \ref{fig4}. We observe that the density increases 
linearly as soon as $\mu\neq 0$. This is not unexpected, since the density 
is proportional to $\sinh \mu$. However, it makes it even more remarkable 
that in the full theory all $\mu$ dependence precisely cancels. We note 
that this is similar to what is expected to occur in phase-quenched QCD 
when $m_\pi/2<\mu< m_N/3$ \cite{Son:2000xc}. Beyond the critical chemical 
potential, the density increases rapidly in the phase-quenched theory as 
well.

We have also computed the average phase factor in the phase-quenched 
theory using real Langevin dynamics. The results are shown in Fig.\ 
\ref{fig5}. For large chemical potential the phase factor goes to zero on 
all lattices, making reweighting impossible. For small chemical potential, 
the phase factor goes to zero exponentially fast as the four-volume 
increases. This is precisely how the average phase factor is expected to 
behave \cite{Splittorff:2006fu}. However, we stress again that in the 
simulations of the full theory no negative impact of the sign problem was 
found.

{\em Summary and outlook --} 
 We have applied stochastic quantization and complex Langevin dynamics to 
study the Silver Blaze problem in the relativistic Bose gas at finite 
chemical potential, both in the full and the phase quenched theory. We 
found precise agreement with theoretical expectations, and no obstacles 
related to the sign problem or in taking the thermodynamic limit. These 
results clearly stimulate the application of this approach to more 
complicated theories with a sign problem, in particular QCD at nonzero 
baryon density. 

In this context two aspects have to be mentioned. First,
due to the complexification, the Langevin dynamics no longer takes place
in SU(3) but instead in SL(3, $\mathbb{C}$). This has been discussed in
detail in Ref.\ \cite{Aarts:2008rr}, where first numerical results for QCD
with a nonzero density of static quarks can be found.
 Second, the inclusion of dynamical fermions, not discussed explicitly in 
Ref.\ \cite{Aarts:2008rr}, is relatively straightforward. After 
integrating out the fermion fields, the fermion determinant contributes to 
the force term for the gauge fields. This is not different from any 
conventional algorithm used in lattice QCD simulations, except that the 
force is now complex, making necessary the extension from SU(3) to 
SL(3,$\mathbb{C}$). For details concerning the inclusion of fermions in 
Langevin dynamics, see e.g.\ Ref.\ \cite{Batrouni:1985jn}. Work in this 
direction is currently in progress.


\begin{acknowledgments}
 Discussions with I.O.\ Stamatescu and S.\ Hands are greatly appreciated. 
This work is supported by STFC.
 \end{acknowledgments}


\end{document}